# Autocatalytic mechanism of pearlite transformation in steel

I.K. Razumov[1,2,*], Yu.N. Gornostyrev[1,2,4], and M.I. Katsnelson[3,4]

[1]Institute of Metal Physics, UB of RAS, 18 S. Kovalevskaya st., Ekaterinburg 620990, Russia
[2]Institute of Quantum Materials Science, 5 Konstruktorov st., Ekaterinburg 620072, Russia
[3]Radboud University, Institute for Molecules and Materials, Heyendaalseweg 135, Nijmegen, 6525 AJ, Netherlands
[4]Ural Federal University, Department of Theoretical Physics and Applied Mathematics, 19 Mira st., Ekaterinburg
620002, Russia

A model of pearlite colony formation in carbon steels with ab-initio parameterization is proposed. The model describes the process of decomposition of austenite and cementite formation through a metastable intermediate structure by taking into account the increase of the magnetic order under the cooling. Autocatalytic mechanism of pearlite colony formation and the conditions for its implementation have been analyzed. We demonstrate that pearlite with lamellar structure is formed by autocatalytic mechanism when thermodynamic equilibrium between the initial phase (austenite) and the products of its decomposition (cementite and ferrite) does not take place. By using model expression for free energy with first-principles parameterization we find conditions of formation of both lamellar and globular structures, in agreement with experiment. The transformation diagram is suggested and different scenarios in the kinetics of decomposition are investigated by phase field simulations.

## I. INTRODUCTION

Pearlite is one of the main structural units of carbon steels which have a significant effect on their properties [1,2]. It is formed by decomposition of austenite ($\gamma$, fcc Fe-C solid solution) into ferrite ($\alpha$-phase, bcc Fe) and cementite (orthorhombic $\theta$-phase, $Fe_3C$) during slow cooling or annealing at temperature $720^0 - 500^0$C. The brightest feature of pearlite is a rather regular lamellar structure in which $\alpha$ and $\theta$ phases are regularly alternated. Pearlite transformation (PT) is an example of eutectoid decomposition which was observed also in many non-ferrous alloys [3–5] below some critical (eutectoid) temperature. Despite numerous studies of PT motivated by its great practical importance for metallurgy, the mechanism of formation of the regular lamellar structure remains unclear.

The proposed theoretical models of PT are focused mostly on the stage of steady-state growth of the pearlite colony and on the problem of stability of the transformation front [6–13]. At the same time, the problems with early stages of the colony formation such as nucleation of cementite remain out of scope of the proposed models. Besides, the mechanism of lamellae multiplication by replication [1,14] or splitting [15], which plays an important role in PT is still under discussion. Also, moving factors of the transition from lamellar to globular pearlite structure with increasing temperature is currently not well understood [16–20].

There is a certain similarity between PT and other diffusion phase transformations which result in the formation of the lamellar structure. One of them is eutectic colony growth which appears behind the front of solidification and is driving by temperature gradient [21,22]. This transformation is determined by fast diffusion at solidification front and/or decomposition of some intermediate states [23,24]. Another example is spinodal decomposition driven by the moving grain boundary (GB) in system with negative mixing energy, $v <0$ [25]. The stability of transformation front is guaranteed automatically in this case and lamellar structure formation is controlled by the redistribution of alloying elements along GB. These observations point out an importance of the acceleration of diffusion at the transformation front as was discussed in relation to the PT problem (see Refs. [6,8,11]). Note, though redistribution of carbon plays a decisive role in the PT the austenite remains stable with respect to the carbon decomposition, $v_\gamma > 0$ [26].

Despite a great practical importance of PT and longtime interest of researchers the mechanisms of this transformation are still poorly understood. In this work we demonstrate that the formation of lamellar structure is a natural part of scenario of PT when the free energy of the system has a special form in which thermodynamic equilibrium between parent $\gamma$-phase and both transformation products ($\alpha$ and $\theta$) is impossible. In this case the pearlite colony can emerge by some kind of autocatalytic mechanism when appearance one of the phases ($\alpha$ or $\theta$) stimulates the nucleation of the other one.

We employ the previously proposed model of phase transformation in iron and steel [27,28] and generalize it taking into account the cementite formation. Following Ref. [29] we also assume that Metastable Intermediate Structure (MIS) exists at $\gamma/\alpha$ interface due to magnetization induced by an adjacent ferrite plate and nucleation of cementite occurs as result of MIS $\rightarrow \theta$ lattice reconstruction when MIS is saturated by carbon. Thus, according to the scenario developing here the PT at undercooling is primarily induced by arising magnetic order in $\alpha$-phase. It is worthwhile to note that the formation of MIS is closely connected to the known fact (see Refs. [30,31]) that the ground state of ferromagnetically ordered $\gamma$-Fe has a strong tetragonal distortion.

## II. METHODS

### A. Effective free energy functional

Here we generalize the previously proposed model [27,28] of $\gamma - \alpha$ transformation by taking into account the cementite formation. In this approach all relevant degrees of freedom (lattice and magnetic) as well as the carbon diffusion redistribution during $\gamma - \alpha$ and $\gamma - \theta$ phase transformations should be included in consideration. We assume that the nucleation of cementite occurs at the $\gamma/\alpha$ interface and $\gamma - \theta$ lattice reconstruction follows the transformation path, which includes the formation of the metastable intermediate structure (MIS) [29].

The pearlite formation is controlled by the carbon diffusion [1,2] which is slow process in contrast with $\gamma - \alpha$ and $\gamma - $ MIS lattice reconstruction carried out by the fast cooperative displacements of Fe atoms. Therefore, we assume that the variables describing the lattice reconstruction take quickly their equilibrium values and the local carbon concentration $c(\mathbf{r})$ remains a single variable which determines the slow evolution.

Since $\alpha$ and $\theta$ phases in pearlite colonies are usually conjugated with small mismatch, whereas the lattice coherency is lost on the transformation front [32], we neglect the elastic energy contribution within the simple model under consideration. Thus, after excluding the fast variables the effective free energy functional can be written in form [33]:

$$F = \int \left( f_{eff}(c,T) + \frac{k_c}{2}(\nabla c)^2 \right) d\mathbf{r} \qquad (1)$$

where $f_{eff}(c,T)$ is effective free energy density for a homogeneous state:

$$f_{eff}(c,T) = \min\{f_\alpha(c,T), f_\gamma(c,T), f_\theta(c,T)\} \qquad (2)$$

and $f_{\gamma(\alpha,\theta)}(c,T)$ is local density of free energy of austenite (ferrite, cementite) at a carbon concentration $c$ and temperature $T$. This means, the phase with lowest energy with a fixed value of local carbon concentration is quickly realized at a given point in space. A similar approach for pearlite free energy was previously used in [11].

To determine the energies $f_{\gamma(\alpha)}(c,T)$ we use the earlier proposed model [28] which takes into account both lattice and magnetic degrees of freedom. According to this model:

$$f^\gamma(c,T) = g_{PM}^\gamma - \int_0^{\tilde{J}_\gamma} Q(\tilde{J}'_\gamma, T) d\tilde{J}'_\gamma - T(s_0 + S_\gamma) \qquad (3)$$

$$f^\alpha(c,T) = g_{PM}^\alpha - \int_0^{\tilde{J}_\alpha} Q(\tilde{J}'_\alpha, T) d\tilde{J}'_\alpha - TS_\alpha \qquad (4)$$

where $s_0$ is the high-temperature limit of the entropy difference between the phases including phonon contribution, $S_{\alpha(\gamma)}$ is configurational entropy of carbon in $\alpha(\gamma)$ phase; $Q(T) \equiv <\mathbf{m}_0 \cdot \mathbf{m}_1> / m^2$ is the spin correlation function dependent on temperature according to Oguchi model [34], $\tilde{J}_{\gamma(\alpha)}(c) = g_{PM}^{\gamma(\alpha)}(c) - g_{FM}^{\gamma(\alpha)}(c)$ is exchange energy,

$$g_{PM(FM)}^\gamma(c) = \tilde{g}_{PM(FM)}^\gamma + \varepsilon_{PM(FM)}^\gamma c + v^\gamma c^2 / 2$$
$$g_{PM(FM)}^\alpha(c) = \tilde{g}_{PM(FM)}^\alpha + \varepsilon_{PM(FM)}^\alpha c + v^\alpha c^2 / 2 \qquad (5)$$

$\tilde{g}_{PM(FM)}^{\gamma(\alpha)}$ are energies of para (ferro)magnetic pure Fe found from the fitting to ab initio computational results [31,35]; $\varepsilon_{PM(FM)}^{\gamma(\alpha)}$ and $v^{\gamma(\alpha)}$ are solution and mixing energies of carbon in fcc (bcc) lattice. During PT, the carbon concentration becomes rather high and reaches the value $c$=0.25. Therefore, the contribution proportional $c^2$ which characterizing carbon-carbon interaction was taken into account in (5). Note that, within this model, a strong temperature dependence of the free energy of $\alpha$-Fe originates rather from the increase of degree of ferromagnetic order during the cooling than from phonon entropy.

We use the traditional lattice-gas model to describe statistical entropy of carbon randomly distributed over interstitials of $\alpha$-Fe. It is a rather good approximation due to very low solubility limit of carbon in $\alpha$ phase, so correlation effects can be neglected. On the other hand, solubility of carbon in $\gamma$ phase is much higher. As in the previous work [28], we assume that carbon atoms may occupy only a part of interstitial positions in $\gamma$ phase. Following Refs. [36,37], we accounted for the repulsive interactions between nearest neighbor's carbon atoms by excluding the part interstitial positions in fcc Fe.

Thus, the configurational entropy of carbon in $\alpha$ and $\gamma$ phase is

$$S_\alpha \approx -kc\ln(c/3)$$
$$S_\gamma = -k[4c\ln(4c) + (1-4c)\ln(1-4c)]/4 \qquad (6)$$

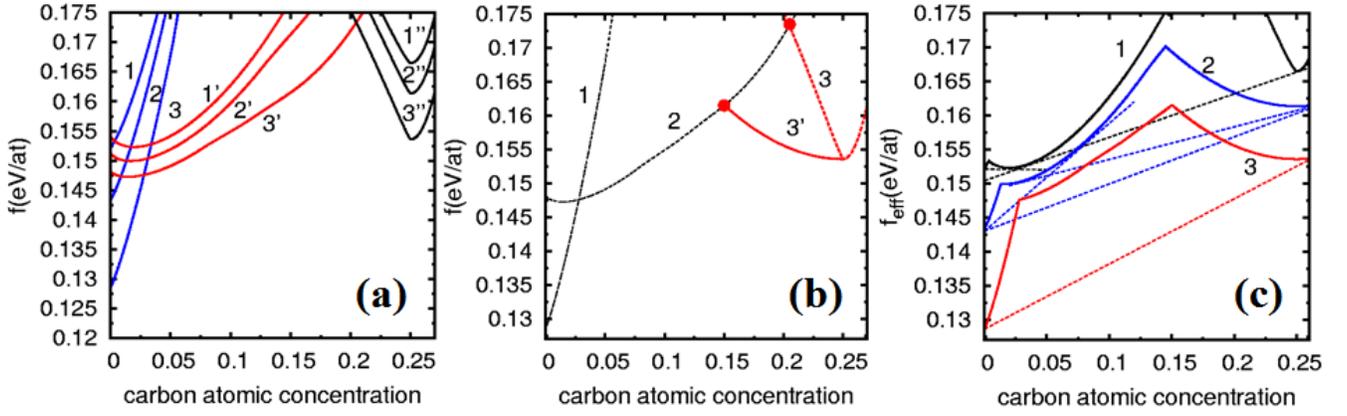

Fig.1. Variants of phase equilibrium in Fe-C system with triple-well thermodynamic potential *f(c)*. **(a)** Calculated density of free energy of $\alpha, \gamma, \theta$ phases in the model with ab initio parameterization at *T*=1050K (1,1',1''), 900K (2,2',2''), 750K (3,3',3''). **(b)** Change in the conditions of cementite formation near the ferrite boundary at 750K; free energy density of $\alpha, \gamma, \theta$-phase (1,2,3) and the effective free energy density of cementite (3'). **(c)** Resulting effective density of free energy as a function of carbon concentration; dotted lines are tangents to the free energies of the phases.

Following Ref. [41] the concentration dependence of the free energy of cementite can be presented as

$$f_\theta(c,T) = f_{\alpha Fe}(T) + \Delta f_{\alpha\theta}(T) + \Delta f_\theta^{(1)}(c,T) + f_{int}(c,T) \quad (7)$$

where $f_{\alpha Fe}(T) = f_\alpha(0,T)$ is the free energy density of pure $\alpha$ iron, $\Delta f_{\alpha\theta}(T)$ is the free energy density of formation of stoichiometric cementite from pure components (bcc Fe and graphite) which is known from CALPHAD [38] or ab initio calculations [39,40], $c_{cem} = 0.25$ is the stoichiometric composition of cementite, $\Delta f_\theta^{(1)}(c,T)$ is the variation of free energy density of cementite due to deviation from stoichiometry calculated in [41] within the model of the regular carbon-vacancy solution, $f_{int}(c,T)$ is an additional contribution to the free energy caused by the interactions of over-stoichiometric carbon atoms. The details of parameterization of the formulas (3)–(7) see in Appendix.

Following the results of Ref. [29] we assume that the cementite nucleation occurs by displacive mechanism in the ferromagnetic region which exists near the ferrite plate and propagates further into the bulk. Herewith, the MIS formed at $\alpha/\gamma$ interface provides the easer and faster realization of $\gamma - \theta$ phase transformation and maintains the lattice coherence. Without describing in detail the process of nucleation, we accept that cementite emerges in the bulk when the local carbon concentration reaches the value of $c \approx 0.20$ (at *T*=0K) [29]. To take into account MIS effect near the ferrite boundary, we replace the concentration dependence $f_\theta^{(1)}(c)$ [41] by an effective one, so that an intersection point of free energies $f_\gamma(c)$ and $f_\theta(c)$ shifts to the left by the value $\Delta c_{bound} \sim 0.05$. We also assume that the carbon concentration *c*=0.25 is reached primarily in the bulk phase with higher carbon solubility, so the energy $f_\theta(c_{cem})$ remains unchanged.

## B. Simulation of transformation kinetics

To study evolution of the microstructure during PT we solved numerically the nonlinear diffusion type equation describing the distribution of carbon $c(\mathbf{r},t)$

$$\frac{\partial c}{\partial t} = -\nabla \mathbf{I}, \quad \mathbf{I} = -\frac{D(c)}{kT} c(1-c) \nabla \left( \frac{\delta F}{\delta c} \right) \quad (8)$$

where *F(c)* is determined by Eq. (1), $D(c)$ is the diffusion coefficient which is supposed to be different in $\gamma$ and $\alpha$ phases. The simulation was performed at the square grid 800x800 with mirror-symmetric boundary conditions [42] by using Runge-Kutta procedure. Such choice of the boundary conditions allows to model the formation of a single isolated pearlite colony in the considered area. To simulate nucleation of colony, we have chosen the initial state as the $\gamma$-phase with a homogeneous carbon concentration and introduced there a small embryo of ferrite or cementite phase.

## III. RESULTS

### A. Transformation diagram

The local density of free energy of each phases $f_{\alpha(\gamma,\theta)}(c)$ calculated for different temperatures by using the model described above (Eqs. (3)-(7)) are shown in Figs.1a,b together with effective local density of free energy $f_{eff}(c) = \min\{f_\alpha(c), f_\gamma(c), f_\theta(c)\}$ (Fig. 1c).

After exclusion of the fast variables, carbon concentration is the only quantity that determines the phase state of the Fe-C. In this case, carbon concentration can be considered as an order parameter, at least in case of ferrite and pearlite decomposition. As one can see from Fig. 1a, at T=750K ferrite and cementite are preferred for $c < 0.027$ at% and $c > 0.20$ at%, respectively; within the interval 0.027 at% $< c < 0.20$ at% austenite is energetically preferable.

The realization of MIS on $\gamma \to \theta$ transformation pathway [29] facilitates the nucleation of cementite near the $\alpha/\gamma$ interface due to magnetization induced by an adjacent ferrite plate. As a result, the $\gamma \to \theta$ transformation starts near the ferrite plate when reaching smaller carbon concentrations (about 15 at %, see Fig.1b) in comparison with bulk. We assume that stoichiometric cementite ($c=0.25$) exists in paramagnetic state and have the same energy in the bulk and near the ferrite plate as well. Thus, we consider an effective $\gamma \to \theta$ transformation pathway, taking into account the nucleation of cementite on the ferrite (curve 3' in Fig.1b).

Above the eutectoid temperature $T_{evtec}$=1000K two-phase equilibria $\gamma + \theta$ and $\gamma + \alpha$ take place (curve 1 in Fig.1c). When temperature decreases, $\gamma$-phase becomes metastable with respect to decomposition on $\theta$ and $\alpha$ phases, wherein a stable equilibrium $\alpha/\theta$ arises (curve 2 in Fig.1c). With further decrease of temperature, below some critical value all metastable equilibria disappear (curve 3, Fig.1c) and only stable two-phase $\alpha + \theta$ state survives.

These changes in equilibrium conditions will result in different scenarios of austenite decomposition. It is convenient to present them using the transformation diagram (Fig.2) proposed earlier in Ref. [28] which takes into account additionally PT. In this diagram the lines $A_3$ and $A_{cm}$ are boundaries of two phase regions $\gamma + \alpha$ and $\gamma + \theta$ respectively, are constructed from the condition of equality of chemical potentials of carbon in the corresponding phases. The lines $T_0$ and $T_1^{bulk}$ correspond to $\gamma/\alpha$ and $\gamma/\theta$ paraequilibrium condition when the free energies of $\gamma$, $\alpha$ or $\gamma$, $\theta$ phases with the same carbon concentration become equal; these lines are coinciding with the temperature dependence of intersection points of free energy densities of considered phases (see Fig.1a). The line $T_1^{bound}$ is obtained by shifting of about 5at% the line $T_1^{bulk}$ and describes the nucleation of cementite near the $\alpha/\gamma$ interface provided by MIS.

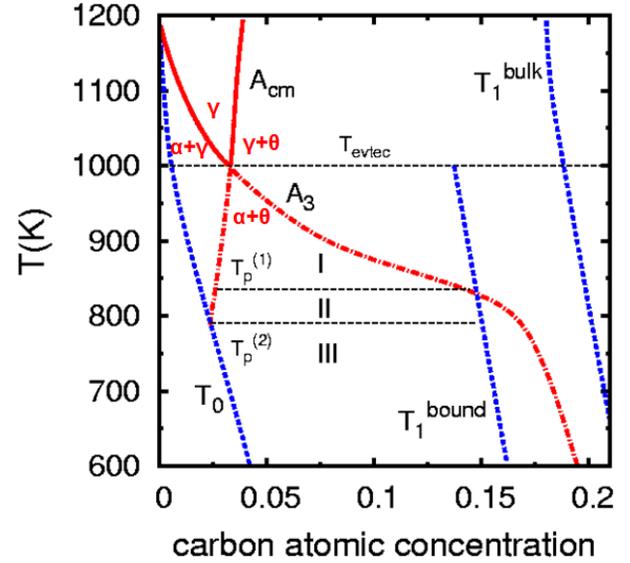

Fig.2. The calculated transformation diagram. The lines $A_3$ and $A_{cm}$ are the boundaries of two-phase regions $\alpha + \gamma$ and $\gamma + \theta$ as well their metastable extensions below the eutectoid temperature $T_{evtec}$; the lines $T_0$ and $T_1$ are lines of instability in respect $\gamma \to \alpha$ and $\gamma \to \theta$ transformation, respectively. The temperature regions I–III are determined by intersection points of these lines.

Below eutectoid temperature $T_{eutec}$ the decomposition $\gamma \to \alpha + \theta$ is possible. As was suggested in Refs. [43,44] the development of PT is expected below $T_{eutec}$ within a window between the metastable extensions of the lines $A_3$ and $A_{cm}$ (corresponding to metastable equilibria $\gamma/\alpha$ and $\gamma/\theta$, respectively) where austenite is supersaturated with respect to both $\alpha$ and $\theta$ phases. In this region the formation one of the phases ($\alpha$ or $\theta$) will stimulate the appearance of another one and therefore results in pearlite colony formation. Here we develop this view and show that this region can be divided into three subdomains I–III where the kinetics of PT is rather different. In the region I both metastable equilibria $\gamma/\alpha$ and $\gamma/\theta$ can be reached (see Fig.1c, curve 2). In the region II only the metastable equilibrium $\gamma/\theta$ survives. Finally, in the region III the metastable equilibria between austenite ($\gamma$) and both transformation products ($\alpha$ and $\theta$) are impossible (Fig.1c, curve 3). It is in the latter case, we can expect the austenite decomposition when appearance one phase ($\alpha$ or $\theta$) will stimulates the fast formation of the another one.

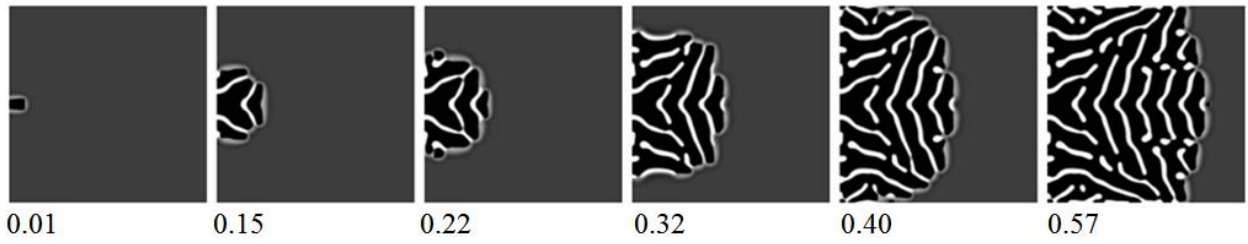

| 0.01 | 0.15 | 0.22 | 0.32 | 0.40 | 0.57 |

Fig.3. Kinetics of lamellar structure growth from a single ferrite nucleus placed on the grain boundary; $T$=675K, $c_0$=0.06; $T$=675K, $c_0$=0.06. The numbers under each fragments correspond to the dimensionless simulation time.

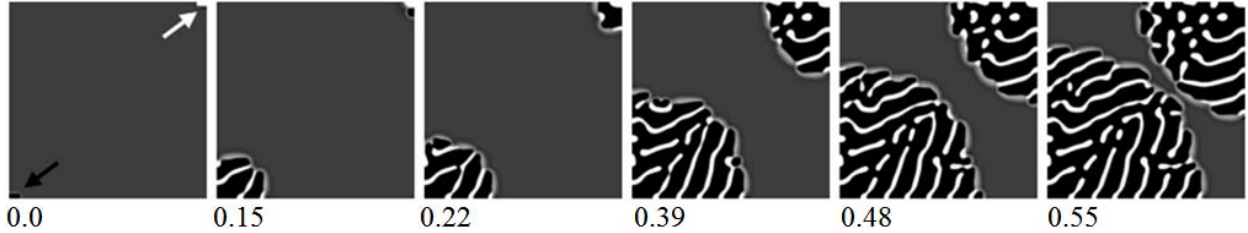

| 0.0 | 0.15 | 0.22 | 0.39 | 0.48 | 0.55 |

Fig.4. Kinetics of lamellar structure growth from a nucleus placed on the grain boundaries junctions (ferrite nucleus in the bottom left and cementite nucleus in the upper right corner are indicated by arrows); $T$=675K, $c_0$=0.06.

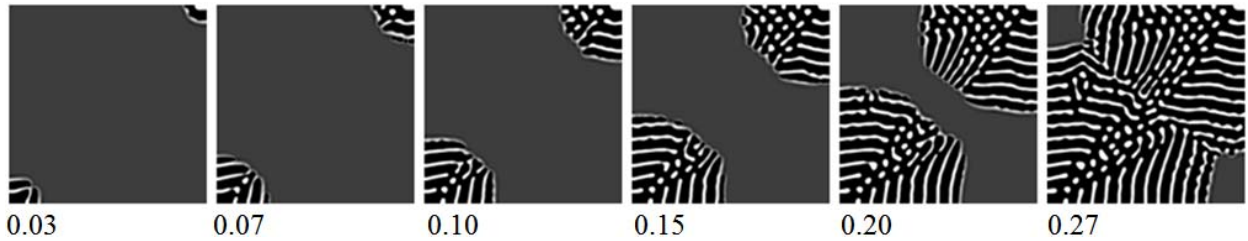

| 0.03 | 0.07 | 0.10 | 0.15 | 0.20 | 0.27 |

Fig.5. Kinetics of lamellar structure growth at shifting the line $T_l$ to the left by $\delta c$ =0.03. The other parameters are the same as in Fig.4.

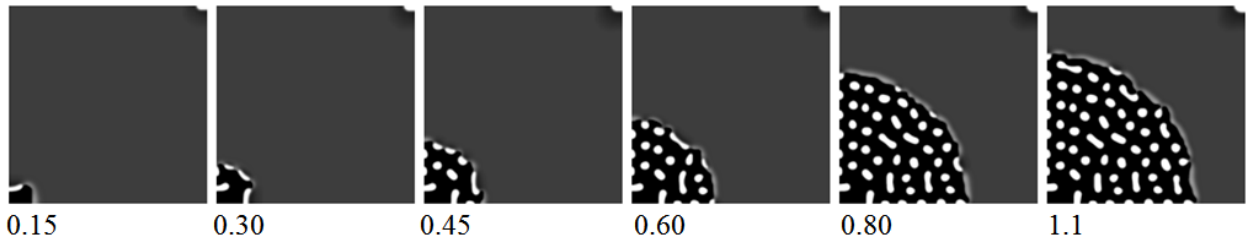

| 0.15 | 0.30 | 0.45 | 0.60 | 0.80 | 1.1 |

Fig.6. Kinetics of globular structure growth; $T$=800K, the other parameters are the same as in Fig.4.

However, the simulation of the decomposition kinetics is required to study microstructure morphology.

## B. Simulation of PT. Lamellar and globular pearlite.

To specify morphology of the transformation product we have carried out phase field simulations of PT starting from homogeneous initial state with a single small ferrite nucleus. We found that below a temperature $T_p^{(1)}$ two scenarios of PT are possible, leading to the formation of either globular (region II) or lamellar structure (region III). The corresponding results are presented in Figs. 3–6. The different levels of carbon concentrations are shown by grayscale, wherein cementite is white and ferrite is black. The time is given in dimensionless units $L^2/D_\alpha$, where the square side is $L$~1mkm (see Appendix).

In the region III, the fine and rather regular lamellar structure is formed regardless of the location of initial embryo (Figs.3,4). In this case, as the first step, carbon is pushed out from embryo of ferrite and its concentration near ferrite interface reaches the threshold value $c(T_1^{bound})$. After this the MIS $\rightarrow \theta$ transformation occurs at interface and carbon flow from $\gamma$-matrix provides saturation of $\theta$-phase and depletion of carbon in surrounding austenite. Since cementite cannot be in equilibrium with $\gamma$-matrix in this region of diagram, the process continues until the critical concentration $c(T_0)$ is reached. After that a new ferrite layer is formed near $\theta$-phase and the process described above is repeated, so the corresponding mechanism can be called autocatalytic. Phase field simulations show that the front movement of the pearlite colony is accompanied by increasing its transverse size. As a result, the pearlite colony gets a fan-type shape in accordance with experiment [1,19]. Herewith, the lamellae do not have a well-marked tendency of normal orientation to the transformation front. At the late stages the space is filled by the domains and the allocation of lamellae is well correlated within each domain. One can assume that the elastic stresses which were not taken into account here, will provide even more regular structures.

Note that a similar pearlite structure can also arise in region III, if we start from one cementite embryo instead ferrite (see Fig.4, upper right corner).

The position of the start line $T_1^{bound}$ of $\gamma \rightarrow \theta$ transformation (see Fig.2) is partly controversial, since the results [29] were obtained at $T$=0K. Therefore, we have performed the calculation at various positions of $T_1^{bound}$. Fig.5 represents the simulation results at choosing the parameters analogous to Fig.4, but the line $T_1^{bound}$ is additionally shifted to the left. It results in the formation of more regular lamellar structure with a smaller interlamellar spacing. Otherwise, the shift of the line $T_1^{bound}$ to the right leads to decreasing of critical temperature of autocatalysis $T_p = \min\{T_p^{(1)}, T_p^{(2)}\}$ and to the coarsening of microstructure; the corresponding kinetics pictures are not presented here.

In the region II PT starts only with ferrite embryos, since they alone can not be in equilibrium with austenite. In this case the condition of autocatalytic multiplication of lamellae is violated and the phase field simulation demonstrates a coarse globular structure (Fig.6). As in the previous case, carbon is pushed out from embryo of ferrite and the chain of transformations $\gamma \rightarrow \text{MIS} \rightarrow \theta$ is realized. However, in this case the line $A_{cm}$ is achieved before the critical concentration $c(T_0)$, so that the metastable phase equilibrium $\gamma / \theta$ is realized, and the new ferritic layer does not appear. As a result, the other scenario of transformation takes place which results in numerous small cementite precipitates in the single ferritic matrix.

In the region I in Fig.2 austenite is decomposed by the conventional nucleation-and-growth mechanism as discussed in Ref. [28] (the corresponding pictures are not shown here). Carbon is pushed out from ferrite embryo and its concentration near ferrite interface reaches the value determined by $A_3$ curve. Since $c(A_3) < c(T_1^{bound})$, the local metastable phase equilibrium $\alpha / \gamma$ is reached, and the formation of cementite does not occur in this case. And vice versa, if we start from one cementite embryo, the local metastable phase equilibrium $\gamma / \theta$ is realized and ferrite does not occur because $c(A_{cm}) > c(T_0)$.

## IV. DISCUSSION

The proposed model based on ab-initio parametrization describes all the most essential features of PT. In particular, the model predicts autocatalytic scenario of quite regular pearlite colonies formation and change in pearlite morphology from lamellar to globular when temperature increase above some critical value $T_p^{(2)}$. It is in agreement with the experimental observations of globular and lamellar pearlite transformations [16–19] and the former take place at smaller undercooling temperatures. Note that the free energies of each phase depend on steel composition and alloying will affect the position of regions I and II.

The suggested mechanism of the autocatalytic PT has some similarity to the spinodal decomposition (SD) of alloys but has also essentially new features. Namely, the austenite remains stable with respect to small fluctuations of carbon concentration and only the formation of ferrite plates stimulates the local saturation of carbon and cementite nucleation. Note, that autocatalytic decomposition of some metastable phases was earlier considered in Refs. [45,11].

Within the approach under consideration, the instability of austenite with respect to $\gamma \rightarrow \alpha + \theta$ decomposition develops stepwise with the temperature decrease. Existence of the threshold temperature of autocatalysis is consistent with the available experimental data [46,47]. For example, according to Ref. [47] the pearlite nucleation rate (in contrast to the growth rate) is close to zero at $T_p^{\exp} < T < T_{evtec}$ and increases abruptly at $T_p^{\exp} \approx 820$K. Similar behavior was discussed in Ref. [46] where a narrower temperature interval with the close to zero nucleation rate was observed.

An important element of the proposed model is the assumption that nucleation of cementite is facilitated near the $\alpha / \gamma$ interface due to magnetization induced by an adjacent ferrite plate. We assume that the Metastable Intermediate Structure (MIS) emerging at $\alpha / \gamma$ interface due to increasing of magnetic order in $\alpha$-phase, plays a decisive role in the formation of cementite and provides the fast lattice reconstruction during the autocatalytic PT.

The model is rather simple and do not take into account real geometry of conjugation $\alpha$ and $\theta$ phases as well as elastic strain due to lattice mismatch. Nevertheless, this approach allows us to construct a realistic transformation diagram and investigate the kinetics of pearlite colonies formation. The results may also be important to eutectic or eutectoid growth of colonies in other systems.

## ACKNOWLEDGMENTS

The research was carried out within the state assignment of FASO of Russia (theme "Magnet" N01201463328), and was partially supported by Act 211 Government of the Russian Federation, contract № 02.A03.21.0006. MIK acknowledges a support from Nederlandse Organisatie voor Wetenschappelijk Onderzoek (NWO) via Spinoza Prize.

## APPENDIX: DETAILS OF PARAMETERIZATION

We use the dissolution and mixing energies of carbon close to those known from experimental data and ab initio calculations: $\varepsilon_{FM}^{\alpha}$ =0.9 [37,47,48], $\varepsilon_{PM}^{\alpha}$ =0.9 [48], $\varepsilon_{PM}^{\gamma}$ =0.3 [50,51], $\varepsilon_{FM}^{\gamma}$ = -0.4 [28,48], $\nu_{\alpha}$ =6 [26], $\nu_{\gamma}$ =1.5 [26,51] (in eV/at). The estimations of $\nu_{\gamma}$ are vary greatly from 1 to 3 eV/at; the values $\varepsilon_{PM}^{\alpha}$, $\varepsilon_{PM}^{\gamma}$ are known only from ab initio calculations; and the difference between ab initio and experimental estimations of $\varepsilon_{FM}^{\alpha}$, $\varepsilon_{PM}^{\alpha}$ is about 0.2eV/at. Note, $\gamma$-phase is stable with respect to small concentration fluctuations because $\nu_{\gamma}$ >0; and the value of $\nu_{\alpha}$ is not essential since carbon solubility in $\alpha$-phase is very small. We neglect the dependence of energies $\varepsilon_{FM(PM)}^{\gamma(\alpha)}$ on temperature, that is partly compensated by the choice of its changed values (within a pointed out error, 0.2eV/at).

The temperature dependence of the difference between cementite and ferrite free energies $\Delta f_{\alpha\theta}(T)$ was chosen close to CALPHAD and ab initio calculations [38–40], with the additional condition, that the curve $A_{cm}$ passes through the eutectoid point ($c$=0.034, $T$=1000K); $\Delta f_{\alpha\theta}(T)$ =0.109- 0.173 $\tau_c$ +0.078 $\tau_c^2$ (in eV/at), where $\tau_c$ =$T/T_c$, $T_c$=1043K. We also assume the value of $f_{int}(c,T)$ is so large for $c$ >$c_{cem}$ that deviation from stoichiometry in this case can be neglected.

The ratios of diffusion coefficients $D_{\alpha}/D_{\gamma}$, $D_{\gamma}/D_{\theta}$ are $10^2 \div 10^3$ [52,53], thus simulation with realistic diffusion coefficients is impossible, but the qualitative tendencies may be revealed when this ratio is choosing sufficiently large. We are also guided by the argument that relaxation of intermediate cementite to its stable state should have time to occur during the growth of the colony. Ultimately, we used the following coefficients: $D_{\alpha}/D_{\gamma}$=$D_{\gamma}/D_{\theta}$ =10.

The square size $L$ is determined by its ratio to interphase boundary width, characterized by the parameter $k_c$, which is the same at $\gamma/\alpha$, $\gamma/\theta$ and $\alpha/\theta$ boundaries. We choose $k_c^2/(kTL^2) \approx 7*10^{-4}$; in that case according to estimations of surface energy [54] the square side is $L$~1mcm.

*e-mail: rik@imp.uran.ru